\documentclass[amsmath, amssymb, floatfix, reprint,onecolumn, notitlepage, 10pt]{revtex4-1}
\usepackage[T1]{fontenc}
\usepackage[utf8]{inputenc}
\usepackage{microtype,bm,bbm,graphicx,booktabs,times}
\usepackage[usenames,dvipsnames]{xcolor}
\usepackage{setspace}
\selectfont{}

\usepackage{hyperref}
\usepackage{subfiles} 
\hypersetup{colorlinks,allcolors=black,linktoc=all}
\usepackage[capitalize]{cleveref}
\crefformat{section}{Sec.~#2#1#3}

\newcommand{\ie}{{\it i.e.}}

\newcommand{\edit}[1]{{\color{black} #1}}

\begin{document}
\title{Fundamental bounds for scattering from absorptionless electromagnetic structures}
\author{Rahul Trivedi$^{1,2}$}
\email{rtrivedi@stanford.edu}
\author{Guillermo Angeris$^{1,2}$}
\author{Logan Su$^1$}
\author{Stephen Boyd$^2$}
\author{Shanhui Fan$^{1,2}$}
\author{Jelena Vu\v{c}kovi\'c$^{1,2}$}
\affiliation{{$^1$E. L. Ginzton Laboratory, Stanford University, Stanford, CA 94305, USA. \\
	         $^2$Department of Electrical Engineering, Stanford, CA 94305, USA.}}

\date{\today}

\begin{abstract}
The ability to design the scattering properties of electromagnetic structures is of fundamental interest in optical science and engineering. While there has been great practical success applying local optimization methods to electromagnetic device design, it is unclear whether the performance of resulting designs is close to that of the best possible design. This question remains unsettled for absorptionless electromagnetic devices since the absence of material loss makes it difficult to provide provable bounds on their scattering properties. We resolve this problem by providing non-trivial lower bounds on performance metrics that are convex functions of the scattered fields. Our bounding procedure relies on accounting for a constraint on the electric fields inside the device, which can be provably constructed for devices with small footprints or low dielectric constrast. We illustrate our bounding procedure by studying limits on the scattering cross-sections of dielectric and metallic particles in the absence of material losses.
\end{abstract}
\maketitle

Understanding the scattering properties of electromagnetic structures has been a problem of fundamental importance in optical science and engineering. Optimization-based design of electromagnetic devices~\cite{molesky2018inverse} has enabled us to realize device functionalities and performances that are far beyond previously anticipated limits~\cite{su2018inverse,piggott2019inverse,su2018fully,piggott2017fabrication,sapra2019inverse}. However, even with the application of such sophisticated design methodologies, the fundamental constraint of Maxwell's equations makes arbitrary device functionalities unlikely. This has raised the question of how to calculate rigorous bounds on the performance achievable by optical devices within a given footprint or for a certain set of design materials.

Several bounds for various performance metrics of interest have been calculated in the past decade. \edit{In particular, calculating bounds on the absorption, extinction, and scattering cross-sections of subwavelength particles has been a problem of great interest due to their diverse applications in imaging, biomedicine and antenna-design \cite{nie1997probing, aizpurua2003optical, alu2008multifrequency, schuller2009optical}}. There have been several attempts to compute these bounds via channel counting arguments~\cite{mclean1996re, hamam2007coupled, kwon2009optimal, yu2010fundamental}, or material-absorption considerations~\cite{miller2016fundamental}. The $\mathbb{T}$-operator formalism has also been used to provide rigorous bounds on scattering from subwavelength particles~\cite{molesky2020t, venkataram2020fundamental, molesky2020fundamental, molesky2019t}. Careful accounting of the cooperative effects of radiation and absorption in electromagnetic scatterers has been used to compute scattering bounds~\cite{kuang2020maximal}. \edit{While the approaches in refs.~\cite{kwon2009optimal, yu2010fundamental, miller2016fundamental, molesky2020t, venkataram2020fundamental, molesky2020fundamental, molesky2019t, kuang2020maximal} have been very successful in providing useful bounds on absorptive electromagnetic structures, they cannot be straightforwardly applied to absorptionless electromagnetic structures}. Bounds on frequency-averaged performance of absorptionless electromagnetic structures have also been provided based on analytical continuation of Maxwell's equations~\cite{shim2019fundamental}, but these bounds are very loose if single-frequency performance is of interest. Lower bounds on error in the electric fields produced by an absorptionless electromagnetic structure relative to a target electric field have been computed by a direct application of Lagrangian duality~\cite{angeris2019computational}, but this procedure requires the target field to be specified at most points in the design region.

In this letter, we consider scattering from absorptionless electromagnetic devices and lower bound frequency-domain performance metrics that can be expressed as convex functions of the scattered fields. Our bounding procedure builds on the principle of Lagrangian duality~\cite{boyd2004convex, bertsekas1997nonlinear}. While a direct application of Lagrange duality to the resulting design problem gives trivial bounds, we show that adding a constraint on the norm of the field inside the electromagnetic device resolves this issue. For low-contrast or subwavelength scatterers, we construct such a constraint from Maxwell's equations and use it to compute bounds on the performance of the device. As an application of this bounding procedure, we use it to calculate upper limits on the scattering cross-section of a 2D absorptionless electromagnetic scatterer.

\begin{figure}[b]
\centering
\includegraphics[scale=0.35]{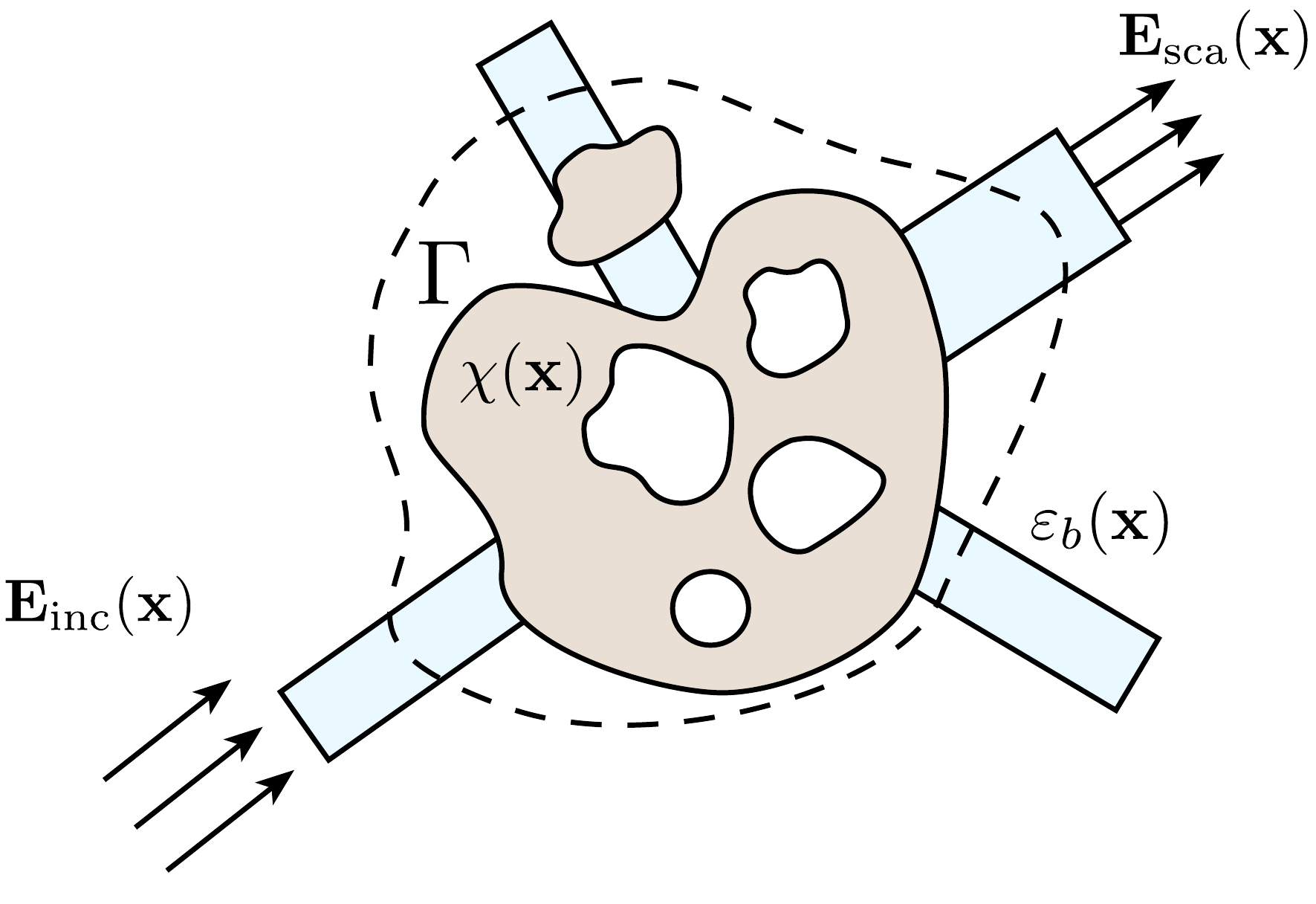}
\caption{\textbf{Schematic}: An electromagnetic device within the region $\Gamma$ with contrast $\chi(\textbf{x})$ embedded in an electromagnetic background with permittivity distribution $\varepsilon_b(\textbf{x})$ is excited with an incident electromagnetic field $\textbf{E}_\text{inc}(\textbf{x})$ to produce a scattered field $\textbf{E}_\text{sca}(\textbf{x})$. }
\label{fig:schematic}
\end{figure}
The setup we consider is shown in Fig.~\ref{fig:schematic}: a lossless electromagnetic device in a design region $\Gamma$ is embedded in a background structure of permittivity distribution $\varepsilon_b(\textbf{x})$. The composition and geometry of the electromagnetic device is described by its contrast $\chi(\textbf{x})$ relative to the background permittivity distribution, \ie, the permittivity distribution inside $\Gamma$ is given by $\varepsilon(\textbf{x}) = \varepsilon_b(\textbf{x}) + \chi(\textbf{x})$. Under excitation by an incident field $\textbf{E}_\text{inc}(\textbf{x})$ propagating in the background medium $\varepsilon_b(\textbf{x})$, the electric field $\textbf{E}(\textbf{x})$ inside the design region $\Gamma$ can be computed from
\begin{align}\label{eq:vol_int_eq}
\textbf{E}(\textbf{x}) = \textbf{E}_\text{inc}(\textbf{x}) + \hat{\textrm{G}}_b \boldsymbol{\Phi}(\textbf{x}), \ \forall \textbf{x}\in \Gamma,
\end{align}
where $\hat{\textrm{G}}_b$ is the Green's function of the background permittivity distribution and $\boldsymbol{\Phi}(\textbf{x}) = \chi(\textbf{x})\textbf{E}(\textbf{x})$ is the polarization current inside the design region. The fields scattered from the device, $\textbf{E}_\text{sca}(\textbf{x})$, are the fields radiated by the polarization current $\boldsymbol{\Phi}(\textbf{x})$. Throughout this letter, except for the scattered fields $\textbf{E}_\text{sca}(\textbf{x})$, all vector fields are only defined with the design region, $\Gamma$. Furthermore, we will assume the following definition of the inner-product $\langle \cdot, \cdot\rangle$ of two vector fields, $\textbf{V}(\textbf{x}), \textbf{U}(\textbf{x})$ defined within $\Gamma$:
\begin{align}
\langle \textbf{V}, \textbf{U}\rangle = \int_\Gamma \textbf{V}^*(\textbf{x})\cdot \textbf{U}(\textbf{x})\,d^3\textbf{x},
\end{align}
with the norm of a vector field $\textbf{V}(\textbf{x})$ being induced by the inner product in the usual way: $\|\textbf{V}\| = \sqrt{\langle \textbf{V}, \textbf{V}\rangle}$.

In a typical optical design problem, we wish to optimize a performance metric (e.g.~transmission through an output port of the device, or the scattering cross-section of the device) with respect to the contrast within the design region. Optimization of many such performance metrics can be mapped to minimization of convex functions of $\textbf{E}_\text{sca}(\textbf{x})$ and consequently as convex functions of $\boldsymbol{\Phi}(\textbf{x})$ since $\textbf{E}_\text{sca}(\textbf{x})$ is linear in $\boldsymbol{\Phi}(\textbf{x})$. \edit{Assuming that the contrast is restricted to vary between two specified limits $\chi_-$ and $\chi_+$, the optimal contrast $\chi_\text{opt}(\textbf{x})$, its electric field $\textbf{E}_\text{opt}(\textbf{x})$, polarization current $\boldsymbol{\Phi}_\text{opt}(\textbf{x})$ and performance $f_\text{opt}$ can be obtained by solving the following optimization problem:
\begin{equation}\label{eq:global_problem}
\begin{aligned}
& \underset{\substack{\boldsymbol{\Phi}, \textbf{E}, \chi \in [\chi_-, \chi_+]}}{\text{minimize}} & & f[\boldsymbol{\Phi}]\\
& \text{subject to} \quad & &\textbf{E}(\textbf{x})  = \textbf{E}_\text{inc}(\textbf{x}) + \hat{\textrm{G}}_b \boldsymbol{\Phi}(\textbf{x}),\quad \forall\textbf{x}\in \Gamma\\
  				      &&&  \boldsymbol{\Phi}(\textbf{x}) = \chi(\textbf{x})\textbf{E}(\textbf{x}), \quad \forall\textbf{x} \in \Gamma,\\
\end{aligned}
\end{equation}
where $f$ captures the performance metric. This nonconvex optimization problem can only be solved locally making it hard to exactly calculate $f_\text{opt}$. A lower bound on $f_\text{opt}$ would provide an estimate of the device performance for given design region $\Gamma$ and contrast limits $\chi_\pm$. One approach to lower bound such a nonconvex optimization problem is to use Lagrange duality~\cite{boyd2004convex, bertsekas1997nonlinear} which constructs a convex, and thus globally solvable, optimization problem that lower bounds the original nonconvex problem. The first step in the application of Lagrangian duality is to construct the Lagrangian $\mathcal{L}$ by adding the constraints in problem \ref{eq:global_problem} to the objective function:
\begin{align}\label{eq:duality}
\mathcal{L}[\boldsymbol{\Phi}, \textbf{E},\chi; \textbf{V}, \textbf{S}]  = f[\boldsymbol{\Phi}] + 2\text{Re}\big[\langle \textbf{V}, \textbf{E} - \textbf{E}_\text{inc} - \hat{\mathrm{G}}_b \boldsymbol{\Phi} \rangle \big] + 2\text{Re}\big[\langle \textbf{S}, \boldsymbol{\Phi}-\chi \textbf{E}\rangle \big].
\end{align}
Here we have introduced vector fields $\textbf{V}(\textbf{x})$ and $\textbf{S}(\textbf{x})$ defined within the design region $\Gamma$, often referred to as the \emph{dual variables}, corresponding to the constraints $\textbf{E}(\textbf{x}) = \textbf{E}_\text{inc}(\textbf{x}) + \hat{\mathrm{G}}_b \boldsymbol{\Phi}(\textbf{x})$ and $\boldsymbol{\Phi}(\textbf{x}) = \chi(\textbf{x})\textbf{E}(\textbf{x})$ respectively. Since $(\boldsymbol{\Phi}_\text{opt}, \textbf{E}_\text{opt}, \chi_\text{opt})$ satisfy the constraints in problem \ref{eq:global_problem}, it follows from Eq.~\ref{eq:duality} that
\begin{align}\label{eq:dual_with_primal}
\mathcal{L}[\boldsymbol{\Phi}_\text{opt}, \textbf{E}_\text{opt},\chi_\text{opt}; \textbf{V}, \textbf{S}] = f[\boldsymbol{\Phi}_\text{opt}] =  f_\text{opt}.
\end{align}
The \emph{dual function} $g[\textbf{V}, \textbf{S}]$ is defined as:
\begin{align}\label{eq:def_dual}
g[\textbf{V}, \textbf{S}] = \inf_{\boldsymbol{\Phi}, \textbf{E}, \chi \in [\chi_-, \chi_+]} \mathcal{L}[\boldsymbol{\Phi}, \textbf{E}, \chi; \textbf{V}, \textbf{S}].
\end{align}
We note that while constructing $g[\textbf{V}, \textbf{S}]$, we minimize $\mathcal{L}$ over all possible values of $\boldsymbol{\Phi}, \textbf{E}$ and $\chi \in [\chi_-, \chi_+]$ instead of only those that satisfy the constraints in problem~\ref{eq:global_problem}. Since the set of all possible $(\boldsymbol{\Phi}, \textbf{E}, \chi)$ also includes $(\boldsymbol{\Phi}_\text{opt}, \textbf{E}_\text{opt}, \chi_\text{opt})$, it immediately follows from Eqs.~\ref{eq:dual_with_primal} and \ref{eq:def_dual} that $g[\textbf{V}, \textbf{S}] \leq f_\text{opt} \ \forall \ \textbf{V}, \textbf{S}$. The best lower bound that $g[\textbf{V}, \textbf{S}]$ can provide is obtained by maximizing it with respect to the dual variables $\textbf{V}$ and $\textbf{S}$. It can be shown that maximizing $g[\textbf{V}, \textbf{S}]$ is a convex optimization problem despite the original problem \ref{eq:global_problem} being nonconvex~\cite{boyd2004convex}. The bound thus obtained is given by (details in the supplement):
\begin{align}
\sup_{\textbf{V}, \textbf{S}} g[\textbf{V}, \textbf{S}] = \min_{\boldsymbol{\Phi}}f[\boldsymbol{\Phi}].
\end{align}
This bound is simply the minimum value of the performance metric $f$ in problem~\ref{eq:global_problem} without accounting for any of its constraints \ie~using the dual function corresponding to the Lagrangian in Eq.~\ref{eq:duality} results in a trivial bound.}

The key insight to resolving this issue is to note that the fields inside the design region cannot be arbitrarily large for most problems of interest. Therefore, we first consider a restriction of this problem where the norm of the difference between the electric field $\textbf{E}(\textbf{x})$ and a reference field $\textbf{E}_\text{ref}(\textbf{x})$ is constrained to be:
\begin{align}\label{eq:norm_const_problem}
\|\textbf{E} - \textbf{E}_\text{ref}\| \leq \alpha \|\textbf{E}_\text{ref}\|.
\end{align}
Here, $\alpha$ is a dimensionless parameter that controls the magnitudes of the fields inside the design region $\Gamma$. The reference field $\textbf{E}_\text{ref}(\textbf{x})$ can be the electric field for any specific device. The Lagrangian function $\mathcal{L}$ corresponding to problem \ref{eq:global_problem} with the field constraint of Eq.~\ref{eq:norm_const_problem} is given by:
\begin{widetext}
\begin{multline}\label{eq:new_lag}
\mathcal{L}[\boldsymbol{\Phi}, \textbf{E},\chi; \textbf{V}, \textbf{S}, \lambda]  = f[\boldsymbol{\Phi}] + 2\text{Re}\big[\langle \textbf{V}, \textbf{E} - \textbf{E}_\text{inc} - \hat{\mathrm{G}}_b \boldsymbol{\Phi} \rangle \big] + 2\text{Re}\big[\langle \textbf{S}, \boldsymbol{\Phi}-\chi \textbf{E}\rangle \big] + \lambda(||\textbf{E} - \textbf{E}_\text{ref}||^2 - \alpha^2 ||\textbf{E}_\text{ref}||^2),
\end{multline}
\end{widetext}
where, compared to Eq.~\ref{eq:duality}, we have introduced an additional dual variable $\lambda \geq 0$ for the field constraint. The dual function $g[\textbf{V}, \textbf{S}, \lambda]$ for the Lagrangian in Eq.~\ref{eq:new_lag} can then be constructed by minimizing it over $\boldsymbol{\Phi}, \textbf{E}$ and $\chi \in [\chi_-, \chi_+]$. As is shown in the supplement, the optimal dual value $d(\alpha) = \text{sup}_{\textbf{V}, \textbf{S}, \lambda\geq 0} g[\textbf{V}, \textbf{S}, \lambda]$, can be computed by solving the following conic program~\cite{boyd2004convex, bertsekas1997nonlinear}:
\begin{widetext}
\begin{equation}\label{eq:conic_program}
\begin{aligned}
& \underset{\substack{\textbf{V}, \textbf{S}, \beta, \lambda \geq 0}}{\text{maximize}} & & 2\text{Re}[\langle \textbf{V}, \textbf{E}_\text{ref} - \textbf{E}_\text{inc} \rangle]-f^\star[\textbf{S} - \hat{\textrm{G}}_b^\dagger\textbf{V}] - \int_\Gamma \beta(\textbf{x})\,d^3\textbf{x} - \lambda \alpha^2 \|\textbf{E}_\text{inc}\|^2\\
& \text{subject to}  &&\beta(\textbf{x}) \geq \frac{|\textbf{V}(\textbf{x}) - \chi_\pm \textbf{S}(\textbf{x})|^2}{\lambda} + 2\chi_\pm\text{Re}\left[ \textbf{S}^*(\textbf{x})\cdot\textbf{E}_\text{ref}(\textbf{x})\right], \quad \forall \textbf{x}\in \Gamma,
\end{aligned} 
\end{equation}
\end{widetext}
where $f^\star$ is the Fenchel dual of $f$~\cite{bertsekas1997nonlinear}. The solution of the convex problem \ref{eq:conic_program}, $d(\alpha)$, then provides a lower bound on the solution of problem~\ref{eq:global_problem} provided that the electric field inside the device is constrained to satisfy Eq.~\ref{eq:norm_const_problem}. From a physical standpoint, $d(\alpha)$ captures how the scattering properties of the scatterer depend on the maximum allowed field intensity inside the design region.

\edit{Since problem \ref{eq:global_problem} does not explicitly restrict $||\textbf{E} - \textbf{E}_\text{ref}||$, in order to use the solution of problem \ref{eq:conic_program} to obtain a bound on $f_\text{opt}$, it is necessary to choose $\alpha$ such that Eq.~\ref{eq:norm_const_problem} will be satisfied for all feasible fields. The smallest $\alpha$ that satisfies this requirement is the optimal solution of the following problem:
\begin{equation}\label{eq:non_convex_alpha}
\begin{aligned}
& \underset{\textbf{E}, \chi \in [\chi_-, \chi_+]}{\text{maximize}} && \|\textbf{E} - \textbf{E}_\text{ref}\| / \|\textbf{E}_\text{ref}\|\\
& \textrm{subject to} && \textbf{E}(\textbf{x}) = \textbf{E}_\text{inc}(\textbf{x}) + \hat{\textrm{G}}_b \chi(\textbf{x})\textbf{E}(\textbf{x}), \quad \forall \textbf{x} \in \Gamma.
\end{aligned}
\end{equation}
Problem~\ref{eq:non_convex_alpha} is nonconvex and therefore difficult to solve globally. However, it follows from Eq.~\ref{eq:vol_int_eq} that $\alpha_\text{ub}$, defined below, is an upper bound on the solution of problem~\ref{eq:non_convex_alpha} and hence a valid choice for $\alpha$ in Eq.~\ref{eq:norm_const_problem} (details in the supplement):
\begin{align}\label{eq:global_field_constraint}
\alpha_\text{ub} = \begin{cases}
\frac{\|(\hat{\textrm{I}} - \bar{\chi}\hat{\textrm{G}}_b)^{-1}\hat{\textrm{G}}_b\| \delta \chi}{({1 - \delta \chi \|(\hat{\textrm{I}} - \bar{\chi}\hat{\textrm{G}}_b)^{-1}\hat{\textrm{G}}_b\|}) } & \text{if } \delta \chi \|(\hat{\textrm{I}} - \bar{\chi}\hat{\textrm{G}}_b)^{-1}\hat{\textrm{G}}_b\| < 1 \\
\infty  &\text{otherwise},
\end{cases}
\end{align}
where $\bar{\chi} = (\chi_+ + \chi_-) / 2$, $\delta \chi = |\chi_+ - \chi_- | / 2$ and $\textbf{E}_\text{ref}(\textbf{x}) = (\hat{\textrm{I}} - \bar{\chi}\hat{\textrm{G}}_b)^{-1}\textbf{E}_\text{inc}(\textbf{x})$. Therefore, $d(\alpha_\text{ub})$ is a lower bound on the optimal value of the nonconvex design problem~\ref{eq:global_problem}:
\begin{equation}\label{eq:final_problem}
d(\alpha_\text{ub}) \leq f_\text{opt}
\end{equation}
 It can be noted that $\alpha_\text{ub}$, and consequently $d(\alpha_\text{ub})$ depend on the choice of the design region $\Gamma$ through the background Green's function $\hat{\textrm{G}}_b$ and the limits $\chi_\pm$ on the contrast $\chi(\textbf{x})$. Furthermore, we note that $d(\alpha_\text{ub})$ is a non-trivial bound only when the design region $\Gamma$, $\chi_-$, and $\chi_+$ are chosen such that $\delta \chi \|(\hat{\textrm{I}} - \bar{\chi} \hat{\textrm{G}}_b)^{-1} \hat{\textrm{G}}_b\| < 1$. While this is a shortcoming of the procedure presented in this letter, an improved upper bound on the optimal value of problem~\ref{eq:non_convex_alpha} would likely improve the bound $f_\text{opt}$.}
 
As an example of application of the bounding procedure outlined above, we consider computing upper-bounds on the scattering cross-section of a 2D lossless scatterer. For this problem, the function $f$ can be chosen to be negative of the scattering cross-section expressed in terms of the polarization current density $\boldsymbol{\Phi}(\mathbf{x})$:
\begin{align}
f[\boldsymbol{\Phi}] = -2\text{Im}\big[ \langle \textbf{E}_\text{inc}, \boldsymbol{\Phi}\rangle\big].
\end{align}
We point out that with this choice of $f$, the upper bound on the scattering cross-section will be $-d(\alpha_\text{ub})$. Furthermore, while numerically solving problem~\ref{eq:conic_program}, it is necessary to discretize the vector fields ($\textbf{V}(\textbf{x}), \textbf{S}(\textbf{x})$), scalar fields ($\beta(\textbf{x})$) and the Green's function ($\hat{\text{G}}_b$) within the design region $\Gamma$. For this letter, we adopt the pulse-basis and delta-testing functions for the discretization~\cite{peterson1998computational}. Numerical studies of the convergence of the discretized problem are included in the supplement.
\begin{figure*}[htpb]
\centering
\includegraphics[scale=0.31]{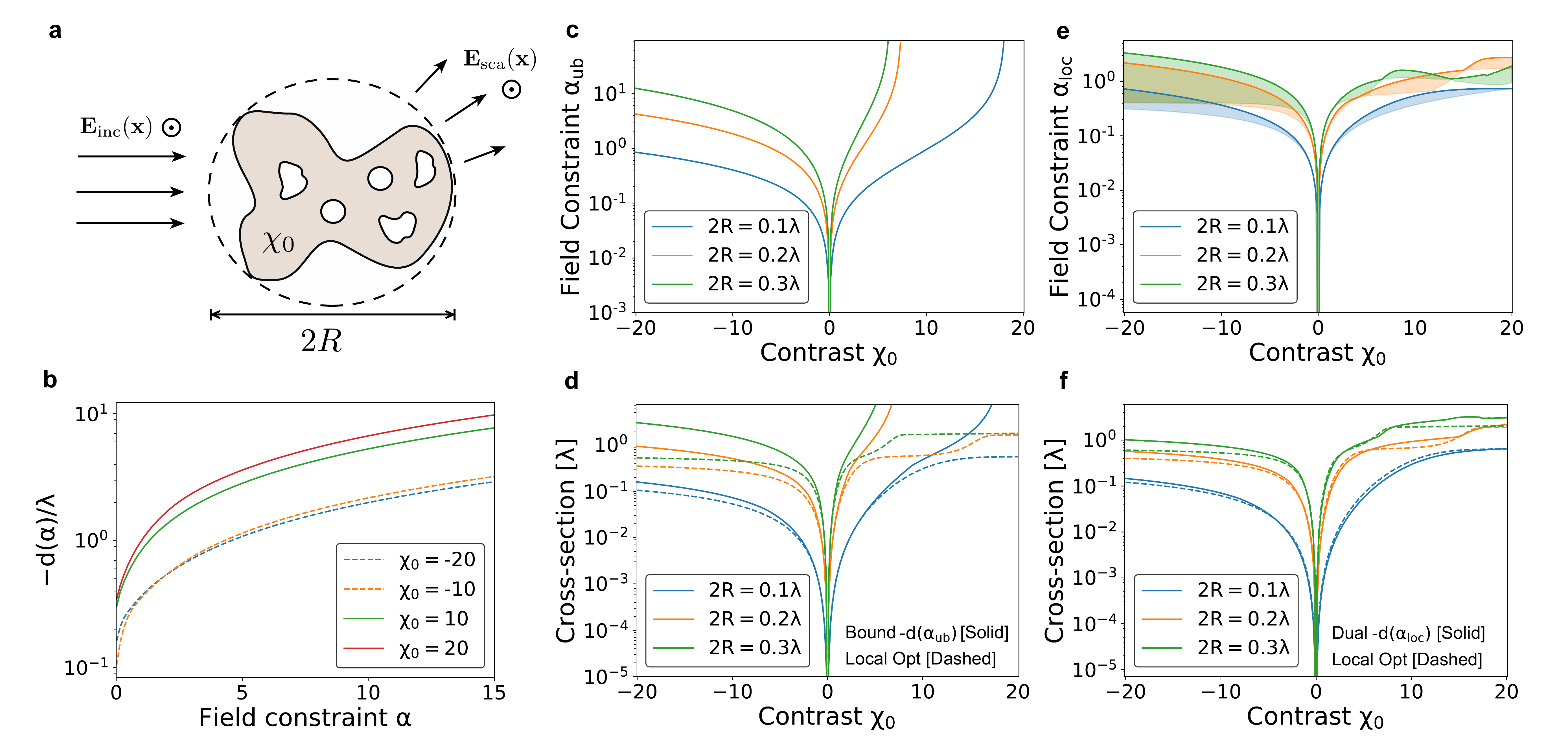}
\caption{\textbf{TE scattering cross-section}: \textbf{a.} Scattering problem schematic. \textbf{b.} The bound $-d(\alpha) / \lambda$ on the scattering cross-section under the field constraint (Eq.~\ref{eq:norm_const_problem}) as a function of $\alpha$ for $2R = 0.2\lambda$. \textbf{c.} The field constraint $\alpha_\text{ub}$ as a function of contrast $\chi_0$ for different $R$. \textbf{d.} The upper bound $-d(\alpha_\text{ub})$ on scattering cross-section. Locally optimized scattering cross-section is shown in dashed lines. \textbf{e.} Local optimum of problem~\ref{eq:non_convex_alpha}, $\alpha_\text{loc}$, as a function of contrast $\chi_0$ for different $R$. The shaded regions indicate the distribution of local optima obtained with 50 different initial conditions. \textbf{f.} The solution of problem~\ref{eq:conic_program} with $\alpha = \alpha_\text{loc}$ shown alongside the locally optimized scattering cross-sections. In all calculations, the vector fields inside $\Gamma$ were represented on a square grid with discretization $\delta x = \lambda / 100$. }
\label{fig:te_scat_cs}
\end{figure*}

We first consider the scattering cross-section for a transverse-electric problem (Fig.~\ref{fig:te_scat_cs}a) where the electric fields are polarized along the $z$-axis while varying spatially with $(x, y)$. We restrict ourselves to a circular design region of radius $R$ with contrast $\chi(\textbf{x})$ varying between $0$ and $\chi_0$ with the background medium being vacuum ($\varepsilon_b(\textbf{x}) = 1$). Fig.~\ref{fig:te_scat_cs}b shows the upper bound on the scattering cross-section under the field constraint Eq.~\ref{eq:norm_const_problem} as a function of $\alpha$, obtained by solving problem~\ref{eq:conic_program} --- our bounds indicate that allowing for higher fields in the design region allows it to have a larger scattering cross-section. Fig.~\ref{fig:te_scat_cs}c shows the field bound $\alpha_\text{ub}$ for the TE scattering problem. For a given radius of the design region, increasing the magnitude of the contrast $\chi_0$ results in an increase in $\alpha_\text{ub}$, with $\alpha_\text{ub}\to\infty$ beyond a cutoff for $\chi_0>0$. Interestingly, the field bound $\alpha_\text{ub}$ does not diverge for negative $\chi_0$. Fig.~\ref{fig:te_scat_cs}d shows the bound on the scattering cross-section obtained by solving problem~\ref{eq:conic_program} with $\alpha = \alpha_\text{ub}$ --- the bound (solid) is compared to the result of local optimization of the scattering cross-section (dashed). We note that for small values of contrast or for small design regions, our bounds are close to the locally optimized results, and significant deviation between the two is only seen due to the divergence in $\alpha_\text{ub}$.

However, an improved constraint on the fields would likely allow us to provide significantly tighter bound for the scattered fields. This is illustrated in Figs.~\ref{fig:te_scat_cs}e and \ref{fig:te_scat_cs}f where we locally solve problem~\ref{eq:non_convex_alpha} to obtain $\alpha_\text{loc}$, which only approximates problem~\ref{eq:non_convex_alpha} and hence doesn't necessarily enforce Eq.~\ref{eq:norm_const_problem} for all feasible fields. As can be seen from Fig.~\ref{fig:te_scat_cs}e, unlike $\alpha_\text{ub}$, $\alpha_\text{loc}$ does not diverge and the corresponding optimal value of $-d(\alpha_\text{loc})$, while not being an actual bound on the scattering-cross section, agrees more closely with the result of locally optimizing the scattering cross-section (Fig.~\ref{fig:te_scat_cs}f).

\begin{figure*}[htpb]
\centering
\includegraphics[scale=0.31]{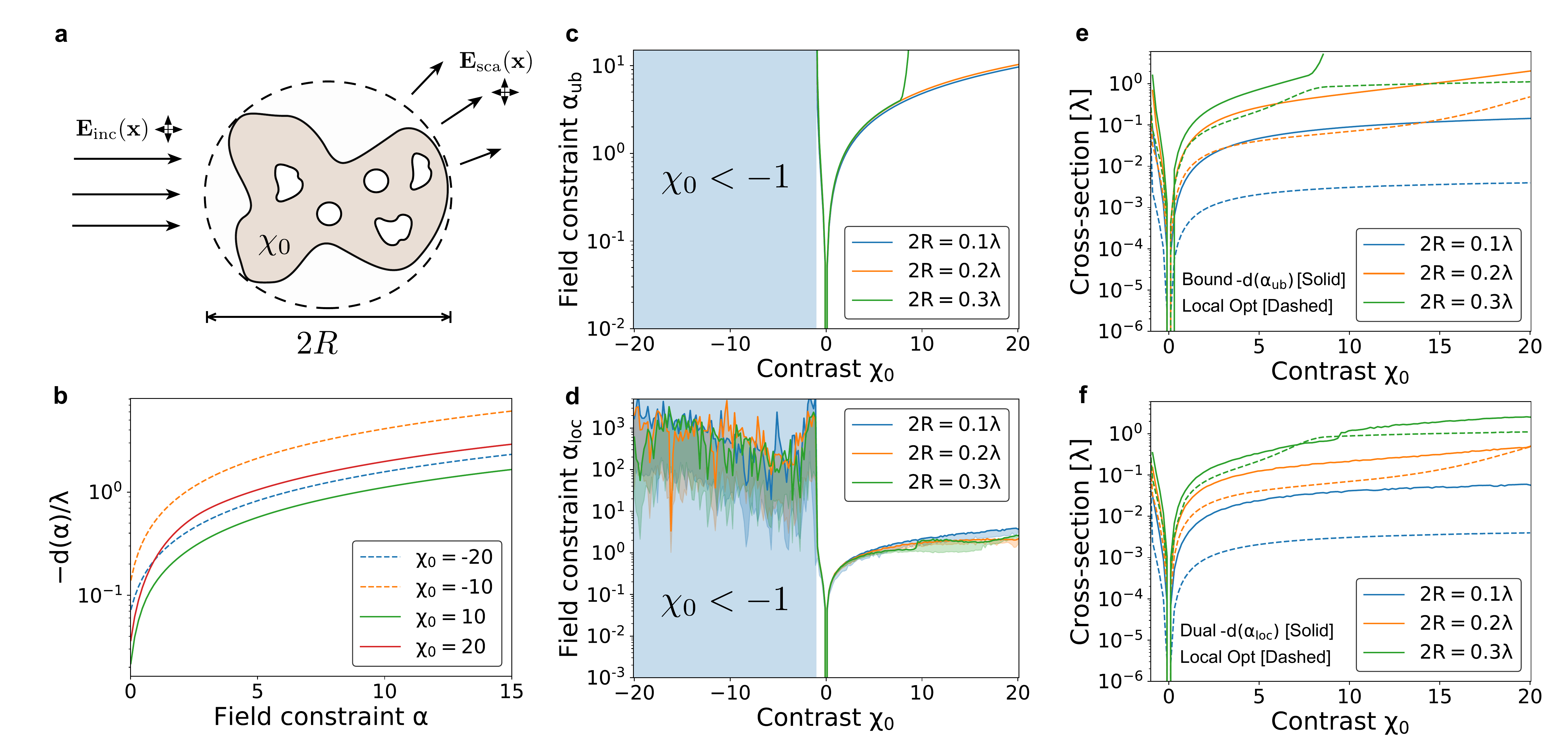}
\caption{\textbf{TM scattering cross-section}: \textbf{a.} Scattering problem schematic. \textbf{b.} The bound $-d(\alpha) / \lambda$ on the scattering cross-section under the field constraint (Eq.~\ref{eq:norm_const_problem}) as a function of $\alpha$ for $2R = 0.2\lambda$. \textbf{c.} The field constraint $\alpha_\text{ub}$ as a function of contrast $\chi_0$ for different $R$. \textbf{d.} Local optimum of problem~\ref{eq:non_convex_alpha}, $\alpha_\text{loc}$, as a function of contrast $\chi_0$ for different $R$. The shaded regions indicate the distribution of local optima obtained with 50 different initial conditions. \textbf{e.} The upper bound $-d(\alpha_\text{ub})$ on scattering cross-section. Locally optimized cross-section is shown in dashed lines. \textbf{f.} The solution of problem~\ref{eq:conic_program} with $\alpha = \alpha_\text{loc}$ shown alongside the locally optimized scattering cross-sections. In all calculations, the vector fields inside $\Gamma$ were represented on a square grid with discretization $\delta x = \lambda / 100$. }\label{fig:tm_scat_cs}
\end{figure*}

Next, we consider the scattering cross-section for a transverse-magnetic problem (Fig.~\ref{fig:tm_scat_cs}a) where the electric fields are polarized in the $xy$-plane, while varying spatially with $(x, y)$. The choice of the design region and the allowed contrast is identical to that of the transverse-electric problem. Fig.~\ref{fig:tm_scat_cs}b shows the solution of the problem~\ref{eq:conic_program} that bounds the scattering cross-section under the field constraint (Eq.~\ref{eq:norm_const_problem}) --- similar to the transverse-electric case, we observe that the scattering cross-section is bounded provided that the fields inside the scatterers are not allowed to be arbitrarily large. Fig.~\ref{fig:tm_scat_cs}c shows $\alpha_\text{ub}$ as a function of the contrast $\chi_0$ --- in contrast to the transverse-magnetic case, we observe that $\alpha_\text{ub} \to \infty$ for $\chi_0 \leq -1$ (\ie, if negative permittivities are allowed in the design region) irrespective of the radius of the design region. For positive $\chi_0$, $\alpha_\text{ub}$ diverges for large $\chi_0$ as indicated in Eq.~\ref{eq:global_field_constraint}. The divergence of $\alpha_\text{ub}$ for positive $\chi_0$ is a consequence of the field bounds (Eq.~\ref{eq:global_field_constraint}) being loose, while the divergence for negative $\chi_0$ is physical. To provide more evidence for this claim, we locally solve the nonconvex optimization problem~\ref{eq:non_convex_alpha} to obtain $\alpha_\text{loc}$ which approximates $\alpha$. Fig.~\ref{fig:tm_scat_cs}d shows $\alpha_\text{loc}$ as a function of the contrast $\chi_0$ for different design region radii $R$ --- we note that for negative $\chi_0$, we obtain extremely large values for $\alpha_\text{loc}$ which are only limited by the spatial discretization used for representing the fields inside the design region $\Gamma$, while for positive values of $\chi_0$ we obtain $\alpha_\text{loc}$ that converges with respect to the spatial discretization (refer to the supplement for numerical studies). Consequently, our bounding procedure suggests that the scattering cross-section for the transverse-magnetic problem is unbounded if negative permittivity materials are allowed in the design region and is bounded if the scatterer is composed entirely of positive permittivity materials (Figs.~\ref{fig:tm_scat_cs}e and \ref{fig:tm_scat_cs}f). This observation is consistent with superscattering effects expected in lossless metallic nanoparticles~\cite{ruan2010superscattering, ruan2011design, mirzaei2014superscattering} due to the existence of surface-plasmon modes with aligned resonant frequencies.

In conclusion, this letter outlined a bounding procedure for absorptionless electromagnetic devices. As an example, we used it to study upper limits on scattering cross-sections of 2D electromagnetic metallic and dielectric scatterers. The generality of the bounding procedure makes it an attractive technique to understand fundamental limits for a variety of electromagnetic design problems. Furthermore, while we have focused on the problem of bounding fields from absorption-less electromagnetic devices, the procedure outlined in this paper can be integrated with the approaches outlined in refs.~\cite{kwon2009optimal, yu2010fundamental, miller2016fundamental, molesky2020t, venkataram2020fundamental, molesky2020fundamental, molesky2019t, kuang2020maximal} to provide tighter bounds for absorptive electromagnetic devices. Finally, we note that the 
general techniques introduced in this paper are not specialized to bounding electromagnetic scattering, but easily extendible to wave-scattering problems in other fields such as accoustics or quantum physics.

\emph{Acknowledgments}: RT acknowledges Kailath Graduate Fellowship. This work has been supported by the AFOSR MURI on attojoule optoelectronics and Samsung. The authors thank Alex Piggott for useful discussions and Alex White, Sattwik Deb Mishra and Geun-Ho Ahn for providing feedback on the manuscript.
\bibliography{library.bib}{}
\end{document}


\title{Supplementary for "Fundamental bounds for scattering from absorptionless electromagnetic structures"}
\author{Rahul Trivedi$^{1}$}
\email{rtrivedi@stanford.edu}
\author{Guillermo Angeris$^{2}$}
\author{Logan Su$^1$}
\author{Stephen Boyd$^2$}
\author{Shanhui Fan$^1$}
\author{Jelena Vu\v{c}kovi\'c$^1$}
\affiliation{{$^1$E. L. Ginzton Laboratory, Stanford University, Stanford, CA 94305, USA. \\
	         $^2$Department of Electrical Engineering, Stanford, CA 94305, USA.}}

\maketitle

\linespread{1.3}\selectfont{}

\section{Dual of the original problem}
Here, we construct the dual function $g(\textbf{V}, \textbf{S})$ corresponding to the Lagrangian in Eq.~4 and show that it provides a trivial bound on the performance metric $f[\boldsymbol{\Phi}]$ in problem~3. From Eq.~4, it follows that:
\begin{align}\label{eq:min_lag_phi_orig}
\inf_{\boldsymbol{\Phi}} \mathcal{L} = -f^\star[\textbf{S} - \hat{\mathrm{G}}_b^\dagger \textbf{V}] - 2\text{Re}[\langle \textbf{V}, \textbf{E}_\text{inc}\rangle] + 2\text{Re}[\langle \textbf{V} - \chi \textbf{S}, \textbf{E}\rangle],
\end{align}
where $f^\star$ is the Fenchel dual of $f$ given by~\cite{bertsekas1997nonlinear}
\begin{align}\label{eq:frenchel}
f^\star[\textbf{N}] = \sup_{\boldsymbol{\Phi}} \bigg(2\text{Re}[\langle \textbf{N}, \boldsymbol{\Phi}\rangle] - f[\boldsymbol{\Phi}]\bigg).
\end{align}
Minimizing Eq.~\ref{eq:min_lag_phi_orig} over $\textbf{E}$, we obtain:
\begin{align}\label{eq:min_lag_phi_e_orig}
\inf_{\textbf{E}, \boldsymbol{\Phi}} \mathcal{L} =
\begin{cases}
-f^\star[\textbf{S} - \hat{\mathrm{G}}_b^\dagger \textbf{V}] - 2\text{Re}[\langle \textbf{V}, \textbf{E}_\text{inc}\rangle] & \text{if } \textbf{V}(\textbf{x}) = \chi(\textbf{x})\textbf{S}(\textbf{x}), \ \forall \textbf{x} \in \Gamma \\
-\infty & \text{otherwise}.
\end{cases}
\end{align}
Finally, $g[\textbf{V}, \textbf{S}]$ can be obtained by minimizing Eq.~\ref{eq:min_lag_phi_e_orig} over $\chi \in [\chi_-, \chi_+]$. Note that unless $\textbf{V}(\textbf{x}) = 0$ and $\textbf{S}(\textbf{x}) = 0, \ \forall \textbf{x} \in \Gamma$, $\chi(\textbf{x})$ can always be chosen so as to give $g[\textbf{V}, \textbf{S}] = -\infty$. Therefore:
\begin{align}\label{eq:dual_orig}
g[\textbf{V}, \textbf{S}] = \begin{cases}
-f^\star[0] & \text{if } \textbf{V}(\textbf{x}) = 0 \text{ and } \textbf{S}(\textbf{x}) = 0, \ \forall \textbf{x} \in \Gamma \\
-\infty & \text{otherwise}.
\end{cases}
\end{align}
As described in the main text, $g[\textbf{V}, \textbf{S}]$ is a lower bound on the optimal solution $f_\text{opt}$ of problem 3 for any choice of $\textbf{V}$ and $\textbf{S}$. Consequently, maximizing $g[\textbf{V}, \textbf{S}]$ over $\textbf{V}$ and $\textbf{S}$ gives the best lower bound on $f_\text{opt}$. It immediately follows from Eqs.~\ref{eq:dual_orig} and \ref{eq:frenchel} that this is given by:
\begin{align}\label{eq:sca_cs}
\sup_{\textbf{V}, \textbf{S}} g[\textbf{V}, \textbf{S}] = -f^\star[0] = \inf_{\boldsymbol{\Phi}} f[\boldsymbol{\Phi}].
\end{align}
This bound is simply the smallest value of the performance metric in problem~3 without accounting for its constraints --- it is consequently a trivial bound on $f_\text{opt}$. It can be noted that for the problem of calculating an upper limit on the scattering cross-section that is considered in the main text, Eq.~\ref{eq:sca_cs} indicates that the scattering cross-section is unbounded irrespective of the choice of design-region $\Gamma$ and the contrast limits $\chi_\pm$.
\section{Dual of the field-constrained problem}
Here, we outline a brief derivation of the dual problem of the field constrained problem. Minimizing the Lagrangian in Eq.~9 over $\boldsymbol{\Phi}(\textbf{x})$, we obtain:
\begin{align}\label{eq:lap_inf_phi}
\inf_{\boldsymbol{\Phi}}\mathcal{L} = -f^\star[\textbf{S} - \hat{\textrm{G}}_b^\dagger \textbf{V}] + 2\text{Re}\big[\langle \textbf{V},  \textbf{E} - \textbf{E}_\text{inc}\rangle \big] - 2\text{Re}\big[\langle \textbf{S}, \chi\textbf{E}\rangle\big]
+ \lambda(\|\textbf{E} - \textbf{E}_\text{ref}\|^2 - \alpha^2 \|\textbf{E}_\text{ref}\|^2).
\end{align}
Minimizing Eq.~\ref{eq:lap_inf_phi} over $\textbf{E}(\textbf{x})$, we obtain:
\begin{align}\label{eq:lap_inf_phi_e}
\inf_{\textbf{E}, \boldsymbol{\Phi}}\mathcal{L} = 2\text{Re}\big[\langle \textbf{V}, \textbf{E}_\text{ref} - \textbf{E}_\text{inc}\rangle \big] -f^\star[\textbf{S} - \hat{\textrm{G}}_b^\dagger \textbf{V}]
- \int_\Gamma \frac{|\textbf{V}(\textbf{x}) - \chi(\textbf{x}) \textbf{S}(\textbf{x})|^2}{\lambda}\,d^3\textbf{x}- 2\text{Re}\big[\langle \textbf{S}, \chi\textbf{E}_\text{ref}\rangle \big].
\end{align}
Finally, to calculate the dual function $g[\textbf{V}, \textbf{S}, \lambda]$ by minimizing Eq.~\ref{eq:lap_inf_phi_e} over $\chi \in [\chi_-, \chi_+]$, note that $\inf_{\textbf{E}, \boldsymbol{\Phi}}\mathcal{L}$ is a concave function in $\chi(\textbf{x})$ and is separable at each point in the domain $\Gamma$, so it will acquire its minimum value with respect to $\chi(\textbf{x})$ at either $\chi_-$ or $\chi_+ \ \forall \ \textbf{x}\in \Gamma$. Consequently,
\begin{align}
g[\textbf{V}, \textbf{S}, \lambda] = 2\text{Re}\big[\langle \textbf{V}, \textbf{E}_\text{ref} - \textbf{E}_\text{inc}\rangle \big] -f^\star[\textbf{S} - \hat{\textrm{G}}_b^\dagger \textbf{V}] - \int_\Gamma \beta(\textbf{x})\,d^3\textbf{x},
\end{align}
where
\begin{align}
\beta(\textbf{x}) = \max_{\chi \in \{\chi_+, \chi_-\}}  \bigg[\frac{|\textbf{V}(\textbf{x}) - \chi\textbf{S}(\textbf{x})|^2}{\lambda} - 2\chi\text{Re}\big(\textbf{S}^*(\textbf{x}) \cdot \textbf{E}_\text{ref}(\textbf{x})\big)\bigg].
\end{align}
Since $g[\textbf{V}, \textbf{S}, \lambda]$ provides a bound on $f_\text{opt}$ for any choice of $\textbf{V}, \textbf{S}$ and $\lambda \geq 0$~\cite{boyd2004convex, bertsekas1997nonlinear}, the best lower bound is then obtained by maximizing it over all the dual variables. This immediately yields the convex optimization problem in the main text (problem~10). Except for an explicit introduction of the polarization current $
\boldsymbol{\Phi}(\textbf{x})$ and the field constraint (Eq.~8), the calculation of the dual problem outlined in this section is similar to that presented in~\cite{angeris2019computational}.
\section{Calculation of the field constraint}
\noindent Here we show that if $\textbf{E}_\text{ref}(\textbf{x}) = (\hat{\text{I}} - \bar{\chi}\hat{\textrm{G}}_b)^{-1}\textbf{E}_\text{inc}(\textbf{x})$, then $\alpha_\text{ub}$ provided in Eq.~12 provides an upper bound on $\|\textbf{E} - \textbf{E}_\text{ref}\|$ for all $\textbf{E}(\textbf{x})$ that are solutions of Eq.~1with $\chi_- \leq \chi(\textbf{x}) \leq \chi_+, \ \forall \ \textbf{x}\in\Gamma$.

Note that Eq.~1 can be rewritten to obtain:
\begin{align}\label{eq:vol_int_rewrite}
\textbf{E}(\textbf{x}) - (\hat{\textrm{I}} - \bar{\chi}\hat{\textrm{G}}_b)^{-1}\hat{\textrm{G}}_b (\chi(\textbf{x}) - \bar{\chi}) \textbf{E}(\textbf{x}) = \textbf{E}_\text{ref}(\textbf{x}).
\end{align}
It immediately follows from the reverse triangle inequality and Eq.~\ref{eq:vol_int_rewrite} that:
\begin{align}\label{eq:rev_trian_ineq}
\|\textbf{E}\| - \|(\hat{\textrm{I}} - \bar{\chi}\hat{\textrm{G}}_b)^{-1}\hat{\textrm{G}}_b (\chi - \bar{\chi}) \textbf{E}\| \leq \|\textbf{E}_\text{ref}\|.
\end{align}
Furthermore, $\|(\hat{\textrm{I}} - \bar{\chi}\hat{\textrm{G}}_b)^{-1}\hat{\textrm{G}}_b (\chi - \bar{\chi}) \textbf{E}\| \geq \|(\hat{\textrm{I}} - \bar{\chi}\hat{\textrm{G}}_b)^{-1}\hat{\textrm{G}}_b \|\cdot \|(\chi - \bar{\chi}) \textbf{E}\|$, where $\|(\hat{\textrm{I}} - \bar{\chi}\hat{\textrm{G}}_b)^{-1}\hat{\textrm{G}}_b \|$ is the operator norm of $(\hat{\textrm{I}} - \bar{\chi}\hat{\textrm{G}}_b)^{-1}\hat{\textrm{G}}_b $. Furthermore, since $\chi_- \leq \chi(\textbf{x})\leq \chi_+ \ \forall \ \textbf{x}\in \Gamma$, it follows that $|\chi(\textbf{x}) - \bar{\chi}| \leq \delta \chi \ \forall \ \textbf{x}\in \Gamma$ where $\delta \chi = |\chi_+ - \chi_-|/2$. Consequently, from Eq.~\ref{eq:rev_trian_ineq}, we immediately obtain:
\begin{align}
(1 - \|(\hat{\textrm{I}} - \bar{\chi}\hat{\textrm{G}}_b)^{-1}\hat{\textrm{G}}_b \| \delta\chi ) \|\textbf{E}\| \leq \|\textbf{E}_\text{ref}\|.
\end{align}
Clearly, if $\|(\hat{\textrm{I}} - \bar{\chi}\hat{\textrm{G}}_b)^{-1}\hat{\textrm{G}}_b \| \delta\chi  < 1$, it follows that $\|\textbf{E}\|$ is bounded above by $\|\textbf{E}_\text{ref}\| / (1 - \|(\hat{\textrm{I}} - \bar{\chi}\hat{\textrm{G}}_b)^{-1}\hat{\textrm{G}}_b \| \delta\chi )$. From Eq.~\ref{eq:vol_int_rewrite}, this bound can easily be translated to a bound on $\|\textbf{E} - \textbf{E}_\text{ref}\|$:
\begin{align}
\|\textbf{E} - \textbf{E}_\text{ref}\| = \|(\hat{\textrm{I}} - \bar{\chi}\hat{\textrm{G}}_b)^{-1}\hat{\textrm{G}}_b  (\chi - \bar{\chi}) \textbf{E} \| \leq \frac{\|(\hat{\textrm{I}} - \bar{\chi}\hat{\textrm{G}}_b)^{-1}\hat{\textrm{G}}_b  \|\delta\chi \|\textbf{E}_\text{ref}\|}{1 - \|(\hat{\textrm{I}} - \bar{\chi}\hat{\textrm{G}}_b)^{-1}\hat{\textrm{G}}_b  \|\delta\chi }.
\end{align}
If $\|(\hat{\textrm{I}} - \bar{\chi}\hat{\textrm{G}}_b)^{-1}\hat{\textrm{G}}_b \| \delta\chi  \geq 1$, then the procedure outlined above cannot be used to generate a bound on $\|\textbf{E} - \textbf{E}_\text{ref}\|$.
\begin{figure}[htpb]
\centering
\includegraphics[scale=0.4]{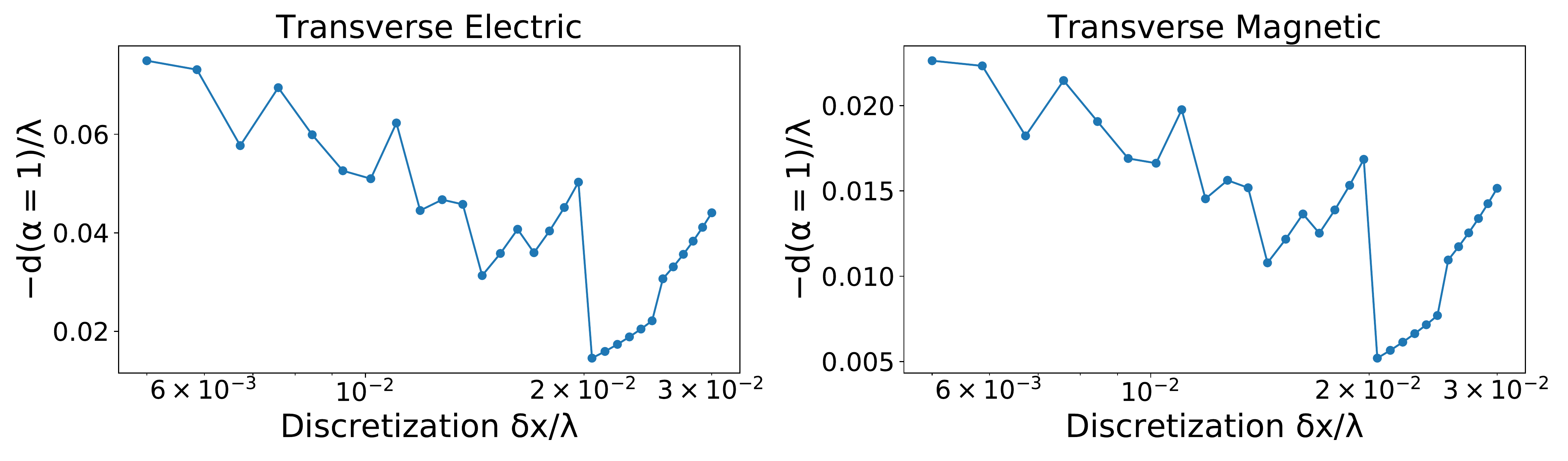}
\caption{Solution of problem 9 with the objective function being the 2D transverse electric and transverse magnetic scattering cross-section for different discretization parameter $\delta x$. The design region is assumed to be a circle of radius $0.1\lambda$ embedded in vacuum and the contrast $\chi_0 = 2.5$.}
\label{fig:conv_conic_program}
\end{figure}

\section{Numerical convergence of the dual problem}
\noindent In order to solve problem~10, all the involved vector fields need to be discretized within the device region $\Gamma$. Furthermore, to ensure that the bounds produced by this problem are not numerical artifacts of the discretization, it is necessary to ensure that the bounds converge with respect to the discretization parameter. This is shown in Fig.~\ref{fig:conv_conic_program} --- the discretization parameter $\delta x$ refers to the length of the pixels used for representing the design region $\Gamma$ for the 2D scattering problems using pulse basis and delta testing functions \cite{peterson1998computational}.

\section{Divergence of field bound}
\begin{figure}[htpb]
\centering
\includegraphics[scale=0.4]{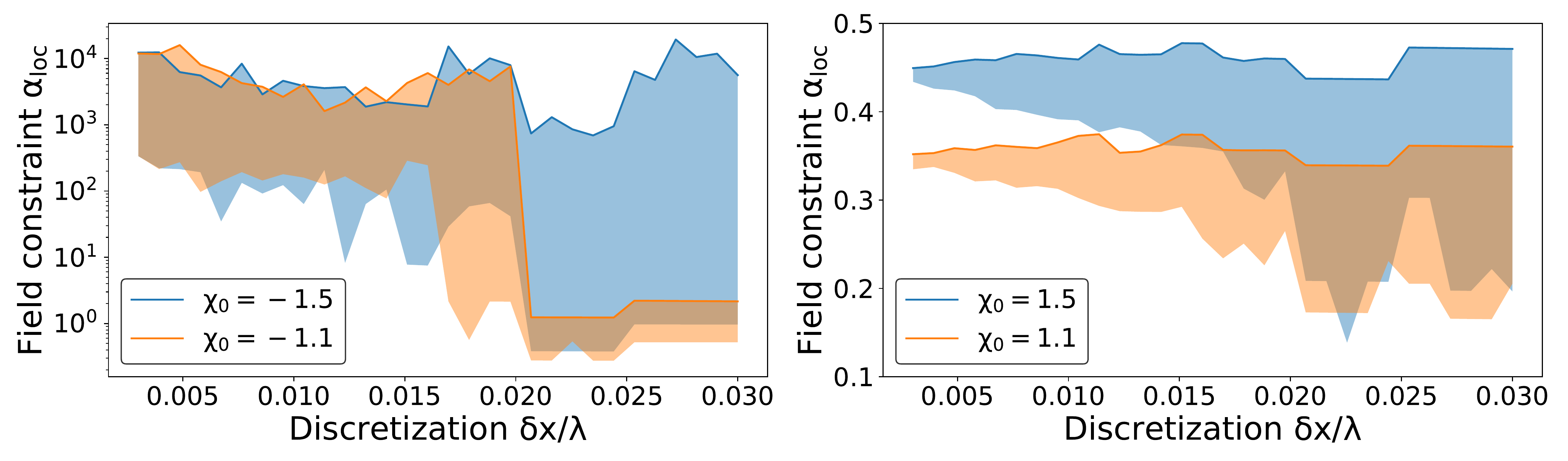}
\caption{The field constraint $\alpha_\text{loc}$ obtained by locally solving problem 11 as a function of the discretization parameter $\delta x$. The shaded regions indicate the range of local-optimas obtained by solving the non-convex problem with 50 different initial conditions. A circular design region with radius $0.1\lambda$ embedded in vacuum is assumed in all calculations.}
\label{fig:alpha_loc_conv}
\end{figure}

\noindent Figure~\ref{fig:alpha_loc_conv} shows locally optimized solutions of problem~11, $\alpha_\text{loc}$, for a transverse magnetic scattering problem as a function of the discretization parameter $\delta x$. We clearly see that for $\chi_0 < -1$, $\alpha_\text{loc}$ does not converge on reducing $\delta x$, while for $\chi_0 > -1$ it converges.

\bibliography{library}{}